\documentstyle[preprint,floats,epsfig,aps]{revtex}
\begin{document}
\def\DESepsf(#1 width #2){\epsfxsize=#2 \epsfbox{#1}}
\input{epsf}

\begin{center}
 \vskip 15mm
{\large CP violation in $B \to PP$ in the SM with
SU(3) symmetry}
 \vskip 15mm
Han-Kuei Fu, Xiao-Gang He, Yu-Kuo Hsiao, and Jian-Qing Shi\\
Physics Department, National Taiwan University, Taipei, Taiwan, R.O.C\\
\vskip 15mm
\end{center}

\begin{abstract}
In this paper we study CP violation in $B\to PP$ decays in the
Standard Model using SU(3) flavor symmetry.
With SU(3) symmetry only seven hadronic
parameters are needed to describe $B\to PP$ decays
in the SM when annihilation
contributions are neglected. The relevant hadronic parameters
can be determined using
known experimental data from $B\to \pi\pi$ and $B\to K \pi$.
We predict branching ratios and CP asymmetries for some of the unmeasured
$B \to PP$ decays. Some of the CP asymmetries can be large and measured
at B factories. The effects of annihilation contributions
can also be studied using present experimental data. Inclusion of
annihilation contributions introduces six more hadronic parameters.
We find that annihilation contributions are in general small,
but can have significant effects on CP asymmetries and some $B_s \to PP$.

\end{abstract}

\newpage
\section{Introduction}
In this paper we study rare charmless hadronic
$B\to PP$ decays
using SU(3) flavor symmetry in the Standard Model (SM). Here $P$ is one of the
SU(3) octet pseudoscalar mesons.
SU(3) analysis for rare charmless B decays have been
studied by many groups and
have obtained several interesting results, such as relations between
different decay branching ratios, and ways to constrain and/or to
determine the phase $\gamma$\cite{1,2,3,4,5,6a}.
SU(3) symmetry for $B\to PP$ decays is expected to be a
good approximation because the energies released in these
decays are larger than the hadronization scale.
Test of SU(3) symmetry has been shown to be possible by using relations
predicted and also using some $B_s\to PP$ decays in an
electroweak model independent way\cite{4}. Here we will take SU(3)
symmetry as our working hypothesis.
We will also study how
SU(3) breaking effects affect the results.

Recently it has been shown that if enough $B\to PP$ decay branching
ratios can be measured, in the frame work of SU(3) symmetry, the
associated hadronic parameters and
their CP conserving final state interaction (FSI) phases, can be
systematically studied\cite{6b}. The CP violating phase $\gamma$ in the KM matrix
can also be constrained. Comparison of the phase $\gamma$ constrained
this way with other constraints, the consistence of the SM
can be checked.
Once the hadronic parameters are determined, CP asymmetries in these
decays can be predicted. We will carry out an analysis
using the most recent data on rare charmless $B\to PP$ decays
to determine hadronic parameters, to predict several
other decay branching ratios and CP asymmetries in $B\to PP$ decays.

We start with a few comments on the determination of the CP violating phase
$\gamma$ using information
from $\epsilon_K$ in $K^0-\bar K^0$ mixing, $B-\bar B$ mixings and $|V_{ub}/V_{cb}|$.
Very stringent constraint on the CP violating phase $\gamma$\cite{6b,8,10} can be
obtained by using experimental information on various KM matrix elements\cite{6b,10}.
Some of the most stringent constraint come from CP violating parameter
$\epsilon_K$, $|V_{ub}/V_{cb}|$, $\Delta m_B$.
The recently measured $\sin(2\beta)$ also provide important information.
Although $B_s -\bar B_s$ mixing has not been measured, one can still
use information on the upper bound on $\Delta m_{B_s}$ to constrain the
phase. One of the method to include $B_s - \bar B_s$ mixing information, in
a global $\chi^2$ fit of $\gamma$, is to use the amplitude
method\cite{ampmd}.
Using the input numerical values of the parameters in these processes
as in Ref. \cite{6b,10}, and the new
averaged value of $\sin(2\beta) = 0.78\pm 0.08$\cite{12,13,14,15},
we obtain the best fit value of $\gamma$ to be $59^\circ$.
The 68\% C.L. and 95\% C.L.
allowed regions, in the Wolfenstein parameters $\rho$ and $\eta$ plane,
are shown for this case in Figure 1.

\begin{figure}[htb]
\centerline{ \DESepsf(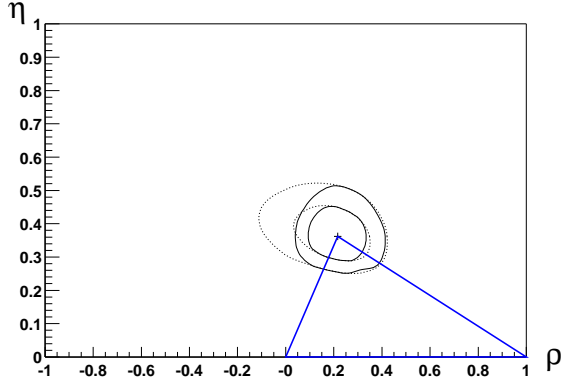 width 8cm)}\label{contour}
\smallskip
\caption{The solid lines are for the fit with $\Delta
m_{B_s}$, and the dashed lines are for the fit without $\Delta
m_{B_s}$
The two regions from smaller to larger corresponds to the
the 68\% C.L. and
the 95\% C.L. allowed regions, respectively.}
\end{figure}

The usage of upper bound on $\Delta m_{B_s}$ to constrain $\gamma$
is not without controversy because the result depends on how the bound is
included. To avoid uncertainties due to this,
we propose to fit the value for $\gamma$ without the use of $\Delta M_{B_s}$
bound. As a by product one can obtain a prediction for the range of
$\Delta m_{B_s}$.
Carrying out an $\chi^2$ analysis, we find that the best values with 68\% C.L.
errors for $\gamma$ and
$\Delta M_{B_s}$, and their
95\% C.L. ranges are given by

\begin{eqnarray}
\gamma &=& 59^{+26^\circ}_{-16^\circ},\;\;\;\;
39^\circ \sim 94^\circ, \mbox{ 95\% C.L. range};
\nonumber\\
\Delta M_{B_s} &=& 17.9^{+4.4}_{-3.9},\;\;\;\; 10.8<\Delta m_{B_s}<26.1,
\mbox{ 95\% C.L. range}.
\end{eqnarray}
In Figure 1. we also show the allowed region (the dashed lines)
in $\rho$ and $\eta$ plane.
 Since experiments constrain $\Delta m_{B_s}$ to be larger than $14.9$ ps$^{-1}$\cite{19} at
the 95\% C.L., the range for $\Delta m_{B_s}$ should be taken to be
$14.9 \sim 26.1$ ps$^{-1}$ at the 95\% C.L..
This prediction is consistent with the prediction of $ \Delta m_{B_s}
= 29^{+16}_{-11}$ ps$^{-1}$\cite{20} from measurement of $\Delta \Gamma_s$
and lattice calculation of $\Delta \Gamma_s/\Delta m_{B_s}$.
The predicted range of $\Delta m_{B_s}$ can be measured at
future hadron colliders, such as LHCb, HERAb and BTeV. This will provide
an important test for the SM.

The value of $\gamma$ obtained above will serve as a reference value for comparison
when we study $B\to PP$ decays. We will make use of the values obtained in two
ways. We will first study the consistence of the value obtained here and the
one to be obtained from $B\to PP$ decays.
The other way is to
fix $\gamma$ at its best fit value determined above and to use experimental data on
$B\to PP$ decays to fix hadronic parameters using SU(3) symmetry, and to predict other
unmeasured branching ratios and CP asymmetries.

In Section II, we will briefly review the SU(3) parameterizations for $B\to PP$ decay
amplitudes in the SM, and
to study the consistence of $\gamma$ by comparing the constraint discussed earlier and that
from $B\to PP$ decays. In Section III we study SU(3) hadronic
parameters, the branching ratios and CP
asymmetries for $B\to PP$ decays. In Section IV, we study
the effects of annihilation amplitudes on $B\to PP$ decays. In Section V, we
discuss some of the implications from our studies and draw conclusions.

\section{SU(3) hadronic parameters and the phase $\gamma$}

In SM the decay amplitudes for $B \to PP$ can be written as
\begin{eqnarray}
A(B\to PP) = <PP|H_{eff}^q|B> = {G_F\over \sqrt{2}}[V_{ub}V^*_{uq} T(q)
+ V_{tb}V^*_{tq}P(q)],
\end{eqnarray}
where $B = (B_u,  B_d,  B_s) = (B^-, \bar B^0, \bar B^0_s)$.
 $T(q)$ contains contributions from
the $tree$ operators as well as $penguin$ operators
due to charm and up
quark loop corrections to the matrix elements,
while $P(q)$ contains contributions purely from
$penguin$ due to top and charm quarks in loops.

SU(3) flavor symmetry can relate different $B\to PP$ decays. Therefore,
knowing some of the branching ratios, other branching ratios and associated
CP violating rate asymmetries can be predicted.
As far as the SU(3) structure is concerned, the effective Hamiltonian
contains $\bar 3$, $6$, and $\overline{15}$ which define three types of
SU(3) invariant amplitudes. We use the notations in Ref. \cite{6b}.
In Table \ref{br}, we list $B\to PP$
decays in terms of the SU(3) invariant amplitudes.

\begin{table}[htb]
\caption{SU(3) decay amplitudes for $B\to PP$ decays.  \label{t1}
} \footnotesize
\begin{eqnarray}
\begin{array}{l}
\hspace{-3mm}
\left.
\begin{array}{l}
\Delta S = 0\\
T^{B_u}_{\pi^-\pi^0}(d) = {8\over \sqrt{2}}C^T_{\overline{15}},\\
T^{B_u}_{\pi^- \eta_8}(d)={2\over \sqrt{6}}
(C^T_{\bar 3}  - C^T_6 + 3 A^T_{\overline {15}} + 3C^T_{\overline {15}}),\\
T^{B_u}_{K^- K^0}(d)=
C^T_{\bar 3}  - C^T_6 + 3 A^T_{\overline {15}} -C^T_{\overline {15}},\\
T^{B_d}_{\pi^+\pi^-}(d) = 2A^T_{\bar 3} +C^T_{\bar 3}
+ C^T_{6} + A^T_{\overline {15} } + 3 C^T_{\overline {15}},\\
T^{B_d}_{\pi^0\pi^0}(d)= {1\over \sqrt{2}} (2A^T_{\bar 3} +C^T_{\bar 3}
+ C^T_{6} + A^T_{\overline {15} } -5 C^T_{\overline {15}}),\\
T^{B_d}_{K^- K^+}(d)= 2(A^T_{\bar 3}  +  A^T_{\overline {15}}),\\
T^{B_d}_{\bar K^0 K^0}(d)= 2A^T_{\bar 3} +
C^T_{\bar 3} - C^T_6 - 3 A^T_{\overline {15}} - C^T_{\overline {15}},\\
T^{B_d}_{\pi^0 \eta_8}(d)= {1\over \sqrt{3}}
(-C^T_{\bar 3}  + C^T_6 + 5 A^T_{\overline {15}} + C^T_{\overline {15}}),\\
T^{B_d}_{\eta_8 \eta_8}(d)={1\over \sqrt{2}}
(2A^T_{\bar 3} + {1\over 3} C^T_{\bar 3} - C^T_6
-A^T_{\overline {15}} + C^T_{\overline {15}}),\\
T^{B_s}_{ K^+ \pi^-}(d) =
C^T_{\bar 3} + C^T_6 -  A^T_{\overline {15}} +3 C^T_{\overline {15}},\\
T^{B_s}_{ K^0 \pi^0}(d) =
-{1\over \sqrt{2}}(C^T_{\bar 3} + C^T_6 -  A^T_{\overline {15}}
-5 C^T_{\overline {15}}),\\
T^{B_s}_{K^0 \eta_8}(d)=
-{1\over \sqrt{6}}(C^T_{\bar 3}  + C^T_6 -  A^T_{\overline {15}}
-5 C^T_{\overline {15}}),
\end{array}
\right.
\left.
\begin{array}{l}
\Delta S = -1\\
T^{B_u}_{\pi^-\bar K^0}(s)= C^T_{\bar 3}
 - C^T_{6} + 3A^T_{\overline {15}} -  C^T_{\overline {15}
 },\\
T^{B_u}_{\pi^0K^-}(s)= {1\over \sqrt{2}} (C^T_{\bar 3}
  - C^T_{6} + 3A^T_{\overline {15} } +7 C^T_{\overline {15}
  })\;,\\
T^{B_u}_{\eta_8K^-}(s)= {1\over\sqrt{6}}(-C^T_{\bar 3}
   + C^T_{6} - 3A^T_{\overline {15}} +9 C^T_{\overline {15}
   }),\\
T^{B_d}_{\pi^+ K^-}(s) =  C^T_{\bar 3}
+ C^T_{6} - A^T_{\overline {15}} + 3 C^T_{\overline {15}
},\\
T^{B_d}_{\pi^0\bar K^0}(s)= -{1\over \sqrt{2}} (C^T_{\bar 3}
+ C^T_{6} - A^T_{\overline {15} } -5 C^T_{\overline {15} }),\\
T^{B_d}_{\eta_8 \bar K^0}(s)=  -{1\over \sqrt{6}} (C^T_{\bar 3}
+ C^T_{6} - A^T_{\overline {15} } -5 C^T_{\overline {15} }),\\
T^{B_s}_{\pi^+\pi^-}(s) = 2(A^T_{\bar 3}
+ A^T_{\overline {15}}),\\
T^{B_s}_{\pi^0\pi^0}(s) = \sqrt{2}(A^T_{\bar 3}
+ A^T_{\overline {15}}),\\
T^{B_s}_{K^+K^-}(s)= 2A^T_{\bar 3} +C^T_{\bar 3}
+ C^T_{6} + A^T_{\overline {15} } + 3 C^T_{\overline {15}
},\\
T^{B_s}_{K^0\bar K^0}(s)= 2A^T_{\bar 3} +C^T_{\bar 3}
- C^T_{6} -3 A^T_{\overline {15} } - C^T_{\overline {15}
},\\
T^{B_s}_{\pi^0\eta_8}(s)= {2\over \sqrt{3}}
( C^T_{6}
+2 A^T_{\overline {15}} - 2C^T_{\overline {15}
}),\\
T^{B_s}_{\eta_8\eta_8}(s)= \sqrt{2}(A^T_{\bar 3} +{2\over 3} C^T_{\bar 3}
- A^T_{\overline {15} } - 2C^T_{\overline {15}}).
\end{array}
\right.
\end{array}
\nonumber
\end{eqnarray}
\label{amp}
\end{table}

\normalsize
In general there are both tree and penguin amplitudes $C^{T,P}_{\bar 3,6,\overline{15}}$,
$A^{T,P}_{\bar 3, 6,\overline{15}}$. $C_{6}$ and
$A_6$ always appear as $C_6 - A_6$ and we
take this combination to be $C_6$. The amplitudes
$A_i$ are referred as annihilation amplitudes. In total there are
10 complex hadronic parameters (20 real parameters with one of them to
be an overall unphysical phase).
However simplification can be made because the following relations in the
SM,

\begin{eqnarray}
C^P_6 &=&
- {3\over 2}
{c_9^{tc} - c_{10}^{tc}\over
c_1-c_2-3(c_9^{uc}-c_{10}^{uc})/2}
C^T_6
\approx
- 0.013
C^T_6
\;,\nonumber\\
C^P_{\overline {15}}(A^P_{\overline {15}})
&=& -{3\over 2}
{c_9^{tc}+c_{10}^{tc}\over c_1+c_2-3(c^{uc}_9+c_{10}^{uc})/2}
C^T_{\overline {15}} (A^T_{\overline {15}})
\approx
+0.015 C^T_{\overline {15}} (A^T_{\overline {15}})
.
\label{P2T}
\end{eqnarray}
Here we have used the Wilson coefficients obtained in
Ref.\cite{dh}.
With the above relations, there are less independent parameters which
we choose them to be,
$C_{\bar 3}^{T,P}(A^{T,P}_{\bar 3})$, $C_6^T$, and $C^T_{\overline {15}}
(A^T_{\overline {15}})$. Using the fact that an overall phase can be removed
without loss of generality, we will set $C^P_{\bar 3}$ to be real, there
are in fact only 13 real independent parameters for $B\to PP$ in the
SM,\\[2 mm]
\hspace*{2 cm}$C_{\bar 3}^P,\;\;C_{\bar 3}^T e^{i\delta_{\bar 3}},\;\;
C^{T}_6e^{i\delta_6},\;\;
C^{T}_{\overline{15}}e^{i\delta_{\overline{15}}}\;\;
A^T_{\bar 3}e^{i\delta_{A^T_{\bar 3}}},\;\;
A^P_{\bar 3} e^{i\delta_{A^P_{\bar 3}}},\;\;
A^T_{\overline{15}} e^{i\delta_{A^T_{\overline{15}}}}.$\\[2 mm]
Further the amplitudes $A_i$ correspond to annihilation contributions and
are expected to be small. In this section, we neglect these amplitudes.
In this case there are
only 7 independent hadronic parameters

\begin{eqnarray}
C_{\bar 3}^P,\;\;C_{\bar 3}^T e^{i\delta_{\bar 3}},\;\;
C^{T}_6e^{i\delta_6},\;\;
C^{T}_{\overline{15}}e^{i\delta_{\overline{15}}}.
\label{ci}
\end{eqnarray}
The phases in the above are defined in such a way that all $C_i^{T,P}$
are real positive numbers.
We will discuss how the annihilation contributions affect the decays in
Section IV.

SU(3) may not be an exact symmetry for $B\to PP$.
The amplitudes $C_i$ for $B\to \pi\pi$ and $B\to K \pi$ will be
different if SU(3) is broken.
At present it is not possible to calculate the  breaking effects. To
have some idea about the size of the SU(3) breaking effects, we
work with the factorization estimate.
To leading order
the relation between the amplitudes for $B\to \pi\pi$ decays
$C_i(\pi\pi)$ and the amplitudes for $B\to K \pi$ decays
$C_i(K \pi)$ can be parameterized as $C_i(K \pi) = r C_i(\pi\pi)$,
and $r$ is approximately given by $r \approx  {f_K\over f_\pi} = 1.22$.

Here we have assumed that
the SU(3) breaking effects in $f_i$ and $F_0^{B\to i}$
are similar in magnitude,
that is, $f_K/f_\pi \approx F^{B\to K}_0/ F^{B\to \pi}_0$.
Using the above to represent SU(3) breaking effect, we can obtain
another set of fitting results.
Compared with $B\to K\pi$, there is also SU(3) breaking effect in
$B_s \to K \pi$ proportional to $F^{B_s \to K}/F^{B\to K}$ or
$F^{B_s \to \pi}/F^{B\to \pi}$. We will take them to be approximately 1.
There are different ways to
determine the hadronic parameters $C_i$ and $\delta_i$. A consistent and systematic
way of carrying out such an analysis is to perform a $\chi^2$ analysis by
taking into account all experimental data on $B\to PP$.
We will use this method  to obtain the hadronic parameters
and also the CP violating phase $\gamma$.

In Table \ref{br} we list present available experimental data on $B\to PP$ decays.
In general the errors for the experimental data in Table \ref{br} are correlated.
Due to the lack of knowledge of the error correlation from experiments, in
our analysis, for simplicity, we take them to be uncorrelated and
assume
the errors obey Gaussian distribution
taking the larger one between $\sigma_+$
and $\sigma_-$ to be on the conservative side. When combining from different
measurements, we take the weighted average.
For the data which only presented as upper bounds,
we assume them to obey Gaussian distribution and taking the
error $\sigma$ accordingly.

We will carry out our $\chi^2$ analysis with the KM matrix elements $V_{us} = \lambda$,
$V_{cb} = A\lambda^2$, $V_{ub} = |V_{ub}| exp(-i\gamma)$     fixed
by\cite{8} $\lambda = 0.2196$,
$A = 0.835$ and $|V_{ub}| = 0.09|V_{cb}|$ and take $\gamma$ to
be a free parameter to be determined in this section. The total parameters to be
determined are the 7 hadronic parameters in Eq. (\ref{ci}) and $\gamma$.

\scriptsize
\begin{table}
\caption{The branching ratios for $B\to PP$ in units of $10^{-6}$.}\label{br}
\begin{center}
\begin{tabular}{|l|l|l|l|l|}
 Branching ratio and  & Cleo&Belle& Babar    &Averaged \\
 CP asymmetries       &\cite{21}  &\cite{22} &\cite{23} & Value \\ \hline \hline
$Br(B_u\to \pi^-\pi^0)$             &    $5.6^{+2.6}_{-2.3}\pm 1.7$                     &    $7.0\pm 2.2\pm 0.8$                            &    $4.1^{+1.1+0.8}_{-1.0-0.7}$                    &         $4.9\pm1.1$\\         \hline
$Br(B_u\to K^-K^0)$             &       $<5.1(90\% \mbox{C.L.})$        &       $<3.8(90\% \mbox{C.L.})$                    &      $<1.3(90\% \mbox{C.L.})$                 &         $0\pm0.8$\\   \hline
$Br(B_d\to \pi^+\pi^-)$         &        $4.3^{+1.6}_{-1.4}\pm 0.5$                 &     $5.1\pm 1.1\pm 0.4$                   &    $5.4 \pm0.7 \pm0.4$                            &            $5.2\pm0.6$\\  \hline
$Br(B_d\to\pi^0\pi^0)$          &        $2.2^{+1.7+0.7}_{-1.3-0.7}$            &   $2.9 \pm 1.5 \pm 0.6$   &    $0.9^{+0.9+0.8}_{-0.7-0.6}$                    &         $1.7\pm0.9$ \\        \hline
$Br(B_d\to K^-K^+)$     &       $<1.9(90\% \mbox{C.L.})$                &       $<0.5(90\% \mbox{C.L.})$                &      $<1.1(90\% \mbox{C.L.})$                     &         $0\pm0.3$\\   \hline
$Br(B_d\to\bar K^0 K^0)$        &       $1.8^{+1.8}_{-1.2} \pm 1.8$             &   $<13(90\% \mbox{C.L.})$   &   $<7.3(90\% \mbox{C.L.})$  &         $1.8\pm2.5$\\ \hline
$Br(B_u\to \pi^-\bar K^0 )$     &    $18.2^{+4.6}_{-4.0}\pm 1.6$                &   $18.8\pm3.0\pm1.5$  &   $17.5^{+1.8}_{-1.7}\pm 1.3$ &         $17.9\pm 1.7$\\   \hline
$Br(B_u\to \pi^0K^-)$           &     $11.6^{+3.0+1.4}_{-2.7-1.3}$                  &    $12.5 \pm 2.4 \pm 1.2$                 &    $11.1^{+1.3}_{-1.2}\pm 1.0$                    &       $11.5\pm 1.3$\\ \hline
$Br(B_d\to \pi^+K^-)$           &    $17.2^{+2.5}_{-2.4}\pm 1.2$                    &    $21.8\pm1.8\pm1.5$                 &    $17.8\pm 1.1 \pm0.8$                       &        $18.6 \pm 1.1$\\   \hline
$Br(B_d\to \pi^0\bar K^0)$          &    $14.6^{+5.9+2.4}_{-5.1-3.3}$                   &    $7.7 \pm 3.2\pm 1.6$               &     $8.2^{+3.1}_{-2.7}\pm1.2$                 &        $8.8\pm 2.3$\\ \hline  \hline
$A_{CP}(B_u\to \pi^-\pi^0)$         &       &   $0.31\pm0.31\pm0.05$    &   $-0.02^{+0.27}_{-0.26}\pm 0.10$ &   $0.13\pm0.21$\\ \hline
$A_{CP}(B_d\to \pi^+\pi^-)$         &                               &   $0.94^{+0.25}_{-0.31}\pm0.09$   &   $-0.02\pm0.29\pm0.07$   &        $0.42\pm0.22$ \\                       \hline
$A_{CP}(B_u\to \pi^-\bar K ^0)$     &         $  0.18\pm0.24$                       &        $0.46\pm0.15\pm0.02$       &        $-0.17\pm0.10\pm0.02$                  &      $0.04\pm0.08$\\  \hline
$A_{CP}(B_u\to \pi^0K^-)$       &         $-0.29\pm0.23$                        &        $-0.04\pm0.19\pm0.03$      &       $0.00\pm0.11\pm0.02$                            &       $-0.05\pm0.09$\\    \hline
$A_{CP}(B_d\to \pi^+K^-)$       &        $-0.04\pm0.16$                             &           $-0.06\pm0.08\pm0.01$   &          $-0.05\pm0.06\pm0.01$                    &       $-0.05\pm0.05$
\end{tabular}
\end{center}
\end{table}

\normalsize

In Figure 2. we show the $\chi^2$ as a function of the phase $\gamma$. From
the figure we see that for the case with exact SU(3) symmetry
$\gamma$ between $20^\circ \sim 160^\circ$,
the $\chi^2$ is reasonably small and allowed at the one sigma level.
Although there are minimal points in the curve, they are not deep enough
to single out one point with high significance. $\gamma$ around $60^\circ$
is certainly allowed. There is no inconsistence between the allowed range
of $\gamma$ obtained here and that in the previous section.
For the case with broken SU(3) symmetry, the region with $\gamma$ around $110^\circ$
is not favored. But $\gamma$ around $60^\circ$ is still allowed at 90\% C.L..
Accurate experimental data in the near future will provide us with more information.

\begin{figure}[htb]
\centerline{ \DESepsf(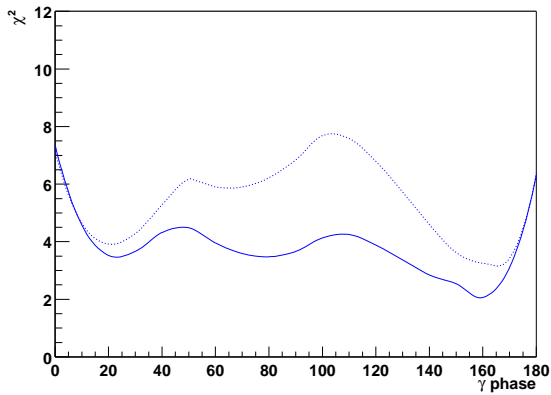 width 8cm)}
\smallskip
\caption {
$\chi^2$ vs. $\gamma$ phase without annihilation terms.
The solid line is for the case with exact SU(3) symmetry,
and the dotted line is for the case with SU(3)
breaking described in the text.} \label{chi}
\end{figure}

\section{Branching ratios and CP asymmetries for $B\to PP$}

In the previous section we have seen that the CP
violating phase $\gamma$ determined using
data from $B\to PP$ and from $\epsilon_K$, $B-\bar B$ mixing and
$|V_{ub}/V_{cb}|$ are not in conflict.
One may want to combine these two to predict
a combined best fit value for $\gamma$. At present the fit from the first section
for $\gamma$ has a much better error range. The combined fit will give a
value for $\gamma$ similar to the one in the previous section\cite{6b}.
In this section we will use
the best fit value of $\gamma= 59^\circ$ from the first section as a known value to study in
more details about $B\to PP$ decays.

The best fit values for the hadronic parameters are given in Table \ref{fp}.
The magnitudes of $C_i$ are the same order of magnitude as the factorization
predictions\cite{6b}. The CP conserving phases $\delta_i$, which can not be calculated
in factorization approximation, can be
determined from the $\chi^2$ analysis performed here. We see from Table \ref{fp} that
these CP conserving phases can be large.

\begin{table}
\caption{The best fit values and their $1\sigma$ errors of the hadronic parameters
using all data in Table \ref{br} with
annihilation terms set to be zero and $\gamma = 59^\circ$.}\label{fp}
\begin{center}
\begin{tabular}{|c|c|c|c|c|}
         &\multicolumn{2}{c}{SU(3) exact}\vline & \multicolumn{2}{c}{ SU(3) break}\vline  \\ \hline
            &   central value   &   error range &   central value   &   error range \\  \hline
$   C^P_{\bar 3}    $   &   0.138   &   0.003   &   0.113   &   0.003   \\  \hline
$   C^T_{\bar 3}    $   &   0.248   &   0.111   &   0.245   &   0.074   \\  \hline
$   C^T_6   $   &   0.155   &   0.112   &   0.142   &   0.092   \\  \hline
$   C^T_{\overline{15}} $   &   0.142   &   0.014   &   0.145   &   0.014   \\  \hline
$   \delta_{  \bar 3}   $   &   $38.10^0$       &        $29.69^0$  &   $34.74^0$   &        $23.51^0$  \\  \hline
$   \delta_ 6   $   &   $83.17^0$       &        $35.97^0$  &   $71.59^0$   &        $29.42^0$  \\  \hline
$   \delta_{\overline{15}}  $   &   $4.78^0$        &       $17.84^0$   &   $3.36^0$    &       $15.31^0$
\end{tabular}
\end{center}
\end{table}

Using the above determined hadronic parameters, one can easily obtain the branching ratios and
CP asymmetries for $B\to PP$.
We use the following definition for the CP violating rate asymmetry,
\begin{eqnarray}
A_{CP} = {\Gamma( B_i \to PP) - \Gamma(\bar B_i \to \bar P \bar P)\over
\Gamma(B_i\to PP) + \Gamma(\bar B_i\to \bar P\bar P)}.
\end{eqnarray}

In general $P$ can be any one of
the SU(3) pseudoscalar octet mesons, $\pi$, $K$ and $\eta_8$. Here we will limit our study
to $P = \pi, K$ to avoid complications associated with $\eta_1$ and $\eta_8$ mixings.
In this case there are total 16 decay modes. Among them
the decay amplitudes for $B_d \to K^-K^+$, $B_s \to \pi^-\pi^+, \pi^0\pi^0$
only receive annihilation contributions.
Since we have neglected annihilation contributions they
would have vanishing branching ratios.
At present none of them have been measured experimentally. The present bound on
$B_d \to K^- K^+$ is consistent with this prediction.

In Tables \ref{BB} and \ref{AA}
we show the results for the branching ratios and CP
asymmetries for the other 13 decays.
We see that the best fit values of the branching ratios for the ones have experimental
measurements are similar and agree with each other within error bars. We also predict
the branching ratios for $B_s \to K^+ \pi^-, K^0 \pi^0, K^-K^+, K^0 \bar K^0$
decays.
These decay modes are predicted to be large and
can be measured at hadron colliders, such as, CDF, HERAb and
LHCb. The SM and SU(3) flavor symmetry can be tested.

SU(3) symmetry predicts some of the CP asymmetries to be equal. From Table \ref{amp}
we obtain

\begin{eqnarray}
&&A_{CP}(B_d \to \bar K^0 K^0) = A_{CP}(B_u \to K^- K^0),\nonumber\\
&&A_{CP}(B_d \to \pi^+ \pi^-) = A_{CP}(B_s \to K^+ \pi^-),\nonumber\\
&&A_{CP}(B_d \to \pi^0 \pi^0) = A_{CP}(B_s \to K^0 \pi^0),\nonumber\\
&&A_{CP}(B_d \to \pi^+ K^-) = A_{CP}(B_s \to K^+ K^-),\nonumber\\
&&A_{CP}(B_u \to \pi^- \bar K^0) = A_{CP}(B_s \to K^0 \bar K^0).
\end{eqnarray}
When SU(3) is broken, in general these relations may no longer hold. However in the
special pattern of the SU(3) breaking  we are dealing with the above relations
still hold. Experimental measurements of CP asymmetries for these modes can
provide important test for the SU(3) flavor symmetry.

In the SU(3) limit there are also some relations between rate differences
defined as,
$\Delta (B_i\to PP) = \Gamma(B_i \to PP) - \Gamma(\bar B_i \to \bar P \bar P)$,
between $\Delta S = 0$ and $\Delta S = -1$ modes due to
a unique feature of the SM in the KM matrix element that\cite{js} $Im(V_{ub}V_{ud}^*V_{tb}^*V_{td})
= - Im(V_{ub}V_{us}^*V_{tb}^*V_{ts})$.
We find\cite{4}

\begin{eqnarray}
&&\Delta (B_d \to \pi^+ \pi^-) = -\Delta (B_d\to \pi^+ K^-),\nonumber\\
&&\Delta (B_d \to \pi^0 \pi^0) = -\Delta (B_d\to \pi^0 \bar K^0),\nonumber\\
&&\Delta (B_d \to \bar K^0 K^0) = -\Delta (B_u\to \pi^- \bar K^0).
\end{eqnarray}
These rate difference relations can also provide important information.

The best fit values for $A_{CP}$ can be large with several of them reaching
30\%, such as the asymmetries for $B_d \to \pi^+ \pi^-, \pi^0\pi^0$, and
$B_s \to K^+ \pi^-, K^0 \pi^0$. $B_d \to \pi^+ \pi^-$ provides the best chance
to measure CP asymmetry. The fact that the size of $A_{CP}$ for these
modes are large, can be easily understood from the following.
Using the above relations, one would obtain

\begin{eqnarray}
&&A_{CP}(B_d \to \pi^+\pi^-) = A_{CP}(B_s \to K^+ \pi^-)
=- A_{CP}(B_d\to \pi^+ K^-) {Br(B_d \to \pi^+ K^-)
\over Br(B_d \to \pi^+ \pi^-)},\nonumber\\
&&A_{CP}(B_d \to \pi^0\pi^0) = A_{CP}(B_s \to K^0 \pi^0)
=- A_{CP}(B_d\to \pi^0\bar K^0) {Br(B_d \to \pi^0 \bar K^0)
\over Br(B_d \to \pi^0 \pi^0)}.
\end{eqnarray}
In all the above cases the ratio of the branching ratios are larger than one, a small
$A_{CP}$ of the decay modes on the right hand side can induce a
large $A_{CP}$ for the decay modes on the left hand side.
The situation with SU(3) breaking case is also similar.

The cases for $B_u \to \pi^- \bar K^0$, $B_d \to K^0 \bar K^0$
and $B_s \to K^0 \bar K^0$, $B_u \to K^- K^0$ are particularly
interesting. In the factorization approximation, the tree amplitude for these modes are
almost zero. In terms of the SU(3) amplitudes that implies
$\Delta C = C^T_{\bar 3} - C^T_6 - C^T_{\overline{15}}$ is close to zero.
CP asymmetries are predicted to be very small.  $\Delta C$ is small, however,
does not follow from SU(3) symmetry. Rescattering effect may make it significantly deviate
from zero. From Table \ref{fp} we
can see that the best fit value for
$\Delta C =^{0.035-0.013\mbox{ $i$... for exact}}_{0.011-0.004\mbox{ $i$... for break}}$
is small, but within errors it can be away from zero. Translating this
into CP violating rate
asymmetries for $B_u \to \bar K^0 \pi^-$ and $B_s \to K^0 \bar K^0$, we see that the best fit
value is small,
but non-zero asymmetries can not be ruled out. This can lead to
large asymmetries for $B_u \to K^- K^0$ and $B_d \to K^0 \bar K^0$ within the error bars,
as can be seen from Table \ref{AA}
experiments.

We note that the CP asymmetry for
$B_u \to \pi^- \pi^0$ is zero in Table \ref{AA} resulting
from SU(3) (or isospin) symmetry.
In principle it
should have a small asymmetry due to the different short distance strong and electroweak
penguins, but it is negligiblly small and have been neglected.

\footnotesize
\begin{table}
\caption{The prediction of the branching ratio
without annihilation terms and $\gamma$=$59^{\circ}$.}\label{BB}
\begin{center}
\begin{tabular}{|c|c|c|c|c|}
Branching ratio &   \multicolumn{2}{c}{ SU(3) Exact}\vline                          &   \multicolumn{2}{c}{ SU(3) break}\vline                          \\  \hline
    &   central value   &   Error(  Max ,   Min )   &   central value   &   Error(  Max ,   Min )   \\  \hline
$B_u \to \pi^- \pi^0$       &   5.3 &   (   6.3 ,   4.2 )   &   5.4 &   (   6.5 ,   4.4 )   \\  \hline
$B_u \to K^- K^0$           &   0.7 &   (   1.1 ,   0.6 )   &   1.1 &   (   1.4 ,   1.0   )   \\  \hline
$B_d \to \pi^+ \pi^-$       &   5.1 &   (   5.7 ,   4.5 )   &   5.0   &   (   5.6 ,   4.4 )   \\  \hline
$B_d \to\pi^0 \pi^0$        &   1.3 &   (   2.1 ,   0.7 )   &   1.1 &   (   1.9 ,   0.6 )   \\  \hline
$B_d \to \bar K^0 K^0$      &   0.7 &   (   1.0   ,   0.6 )   &   1.0   &   (   1.3 ,   0.9 )   \\  \hline
$B_u \to \pi^- \bar K^0$    &   19.1    &   (   20.3    ,   18.0  )   &   19.2    &   (   20.3    ,   18.1    )   \\  \hline
$B_u \to \pi^0 K^-$         &   10.5    &   (   11.1    ,   10.0  )   &   10.8    &   (   11.3    ,   10.3    )   \\  \hline
$B_d \to \pi^+ K^-$         &   18.5    &   (   19.5    ,   17.6    )   &   18.4    &   (   19.3    ,   17.5    )   \\  \hline
$B_d \to \pi^0 \bar K^0$    &   8.6 &   (   9.0   ,   8.1 )   &   8.3 &   (   8.7 ,   7.9 )   \\  \hline  \hline
$B_s \to K^+ \pi^-$         &   4.8 &   (   5.3 ,   4.2 )   &   7.0   &   (   7.8 ,   6.2 )   \\  \hline
$B_s \to K^0 \pi^0$         &   1.2 &   (   2.0   ,   0.7 )   &   1.6 &   (   2.7 ,   0.8 )   \\  \hline
$B_s \to K^+K^- $           &   17.4    &   (   18.3    ,   16.5    )   &   26.6    &   (   27.9    ,   25.3    )   \\  \hline
$B_s \to K^0 \bar K^0$      &   16.8    &   (   17.9    ,   15.8    )   &   25.9    &   (   27.4    ,   24.5    )   \\  \hline
\end{tabular}
\end{center}
\end{table}

\footnotesize
\begin{table}
\caption{The prediction of the CP asymmetry without annihilation
terms and $\gamma$=$59^{\circ}$}\label{AA}
\begin{center}
\begin{tabular}{|c|c|c|c|c|}
Asymmetry   &   \multicolumn{2}{c}{ SU(3) Exact}\vline                          &   \multicolumn{2}{c}{ SU(3) break}\vline                          \\  \hline
    &   Central value   &       Error (Max,Min)             &   Central value   &       Error (Max,Min)             \\  \hline
$B_u \to \pi^- \pi^0$       &   0.00    &   (   0.00   ,   0.00   )   &   0.00   &   (   0.00   ,   0.00   )   \\  \hline
$B_d \to \pi^+ \pi^-$       &   0.32    &   (   0.46    ,   0.18    )   &   0.24    &   (   0.35    ,   0.12    )   \\  \hline
$B_u \to \pi^- \bar K^0$    &   0.00    &   (   0.05    ,   -0.04   )   &   0.00   &   (   0.04    ,   -0.03   )   \\  \hline
$B_u \to \pi^0 K^-$         &   -0.01   &   (   0.06    ,   -0.10    )   &   -0.01   &   (   0.06    ,   -0.10   )   \\  \hline
$B_d \to \pi^+ K^-$         &   -0.09   &   (   -0.05   ,   -0.13   )   &   -0.10    &   (   -0.05   ,   -0.14   )   \\  \hline  \hline
$B_u \to K^- K^0$           &   -0.09   &   (   0.85    ,   -0.91   )   &   -0.03   &   (   0.74    ,   -0.78   )   \\  \hline
$B_d \to\pi^0 \pi^0$        &   0.37    &   (   0.64    ,   -0.58   )   &   0.32    &   (   0.56    ,   -0.38   )   \\  \hline
$B_d \to \bar K^0 K^0$      &   -0.09   &   (   0.85    ,   -0.91   )   &   -0.03   &   (   0.74    ,   -0.78   )   \\  \hline
$B_d \to \pi^0 \bar K^0$    &   -0.06   &   (   0.06    ,   -0.13   )   &   -0.07   &   (   0.06    ,   -0.15   )   \\  \hline
$B_s \to K^+ \pi^-$         &   0.32    &   (   0.46    ,    0.18    )   &   0.24    &   (   0.35    ,   0.12    )   \\  \hline
$B_s \to K^0 \pi^0$         &   0.37    &   (   0.64    ,   -0.58   )   &   0.32    &   (   0.56    ,   -0.38   )   \\  \hline
$B_s \to K^+K^- $           &   -0.09   &   (   -0.05   ,   -0.13   )   &   -0.10    &   (   -0.05   ,   -0.14   )   \\  \hline
$B_s \to K^0 \bar K^0$      &   0.00    &   (   0.05    ,   -0.04   )   &   0.00   &   (   0.04    ,   -0.03   )   \\  \hline
\end{tabular}
\end{center}
\end{table}

\normalsize

At present no CP asymmetry in $B\to PP$ has been measured. To see how sensitive the
bounds on CP asymmetries in Table \ref{br} affect the analysis,
we carried out an analysis using  mostly branching ratio information.
If we do not use any CP violating data, we find that the branching ratios are not affected
very much. However, in this case there is a degeneracy in identifying
particle and anti-particle branching ratios.
This implies that one can only determine the size of the asymmetries
but not the signs. To determine the sign, one should use at least one CP asymmetry
data point to left the degeneracy. For this purpose we select one CP asymmetry data point,
the asymmetry for $B_d \to \pi^+ K^- $, for which all experimental measurements have similar
central values although there is still a large error bar to establish the measurement.
We list the results in Tables
\ref{AAA},  \ref{CCC} and \ref{BBB}.
\begin{table}
\caption{The best fit values and their error ranges for the hadronic parameters
without annihilation terms and $\gamma = 59^\circ$ using
data on branching ratios and CP asymmetry on $B_d \to K^+ \pi^-$.}\label{AAA}
\begin{center}
\begin{tabular}{|c|c|c|c|c|}
\hline
         &\multicolumn{2}{c}{SU(3) exact}\vline & \multicolumn{2}{c}{ SU(3) break}\vline  \\ \hline
            &   central value   &   error range &   central value   &   error range \\  \hline
$   C^P_{\bar 3}    $   &   0.139   &   0.003   &   0.114   &   0.003   \\  \hline
$   C^T_{\bar 3}    $   &   0.280    &   0.112   &   0.271   &   0.074   \\  \hline
$   C^T_6   $   &   0.176   &   0.147   &   0.182   &   0.103   \\  \hline
$   C^T_{\overline{15}} $   &   0.141   &   0.014   &   0.143   &   0.014   \\  \hline
$   \delta_{  \bar 3}   $   &   $29.54^0$       &        $29.86^0$  &   $27.92^0$   &        $19.86^0$  \\  \hline
$   \delta_ 6   $   &   $75.15^0$       &        $32.15^0$  &   $65.18^0$   &        $22.51^0$  \\  \hline
$   \delta_{\overline{15}}  $   &   $-13.33^0$      &        $21.71^0$  &   $-15.50^0$  &        $19.03^0$  \\  \hline
\end{tabular}
\end{center}
\end{table}
From the Table \ref{AAA},
we see that the size of the hadronic parameters $C_i$ are not affected
very much, but the CP conserving phase $\delta_i$ can vary quite a lot, especially for
$\delta_{\overline{15}}$. In terms of the
branching ratios and CP asymmetries we find that branching ratios are similar, but
CP asymmetries can be quiet different which can be seen from Tables
\ref{CCC} and \ref{BBB}. The differences are largely caused by the
differences in $\delta_i$.  It is therefore very important to have good
CP asymmetry measurement which not only provide information for CP violation but also
information for the detailed dynamics of hadronic physics.

\footnotesize
\begin{table}
\caption{The prediction of the branching ratio without
annihilation terms and $\gamma$=$59^{\circ}$ using data on
branching ratios and CP asymmetry in $B_d\to K^+ \pi^-$.}\label{CCC}
\begin{center}
\begin{tabular}{|c|c|c|c|c|}
Branching ratio &   \multicolumn{2}{c}{ SU(3) Exact}\vline                          &   \multicolumn{2}{c}{ SU(3) break}\vline                          \\  \hline
    &   central value   &   Error(  Max ,   Min )   &   central value   &   Error(  Max ,   Min )   \\  \hline
$B_u \to \pi^- \pi^0$       &   5.1 &   (   6.2 ,   4.1 )   &   5.3 &   (   6.4 ,   4.3 )   \\  \hline
$B_u \to K^- K^0$       &   0.8 &   (   1.2 ,   0.7 )   &   1.1 &   (   1.5 ,   1.0   )   \\  \hline
$B_d \to \pi^+ \pi^-$       &   5.1 &   (   5.7 ,   4.6 )   &   5.1 &   (   5.7 ,   4.5 )   \\  \hline
$B_d \to\pi^0 \pi^0$        &   1.6 &   (   2.5 ,   0.8 )   &   1.5 &   (   2.4 ,   0.7 )   \\  \hline
$B_d \to \bar K^0 K^0$      &   0.7 &   (   1.1 ,   0.6 )   &   1.0   &   (   1.4 ,   0.9 )   \\  \hline
$B_u \to \pi^- \bar K^0$    &   19.3    &   (   20.5    ,   18.1    )   &   19.4    &   (   20.6    ,   18.3    )   \\  \hline
$B_u \to \pi^0 K^-$         &   10.6    &   (   11.1    ,   10.0  )   &   10.9    &   (   11.4    ,   10.4    )   \\  \hline
$B_d \to \pi^+ K^-$         &   18.4    &   (   19.3    ,   17.4    )   &   18.1    &   (   19.1    ,   17.2    )   \\  \hline
$B_d \to \pi^0 \bar K^0$    &   8.5 &   (   8.9 ,   8.0 )   &   8.2 &   (   8.7 ,   7.8 )   \\  \hline  \hline
$B_s \to K^+ \pi^-$         &   4.8 &   (   5.4 ,   4.3 )   &   7.1 &   (   7.9 ,   6.3 )   \\  \hline
$B_s \to K^0 \pi^0$         &   1.5 &   (   2.3 ,   0.7 )   &   2.1 &   (   3.3 ,   1.0   )   \\  \hline
$B_s \to K^+K^- $       &   17.3    &   (   17.9    ,   16.4    )   &   26.2    &   (   27.6    ,   24.9    )   \\  \hline
$B_s \to K^0 \bar K^0$      &   17.0  &   (   18.1    ,   16.0  )   &   26.3    &   (   27.9    ,   24.8    )   \\  \hline
\end{tabular}
\end{center}
\end{table}

\footnotesize
\begin{table}
\caption{The prediction of the CP asymmetry without
annihilation terms and $\gamma$=$59^{\circ}$ using
data on branching ratios and CP asymmetry from $B_d \to K^+ \pi^-$.}\label{BBB}
\begin{center}
\begin{tabular}{|c|c|c|c|c|}
Asymmetry   &   \multicolumn{2}{c}{ SU(3) Exact}\vline                          &   \multicolumn{2}{c}{ SU(3) break}\vline                          \\  \hline
    &   Central value   &       Error (Max,Min)             &   Central value   &       Error (Max,Min)             \\  \hline
$B_u \to \pi^- \pi^0$       &   0.00   &   (   0.00   ,   0.00   )   &   0.00   &   (   0.00   ,   0.00   )   \\  \hline
$B_d \to \pi^+ \pi^-$       &   0.21    &   (   0.38    ,   0.03    )   &   0.15    &   (   0.27    ,   0.03    )   \\  \hline
$B_u \to \pi^- \bar K^0$    &   0.00   &   (   0.05    ,   -0.05   )   &   0.00   &   (   0.04    ,   -0.04   )   \\  \hline
$B_u \to \pi^0 K^-$         &   0.06    &   (   0.17    ,   -0.2    )   &   0.09    &   (   0.18    ,   -0.07   )   \\  \hline
$B_d \to \pi^+ K^-$         &   -0.06   &   (   -0.01   ,   -0.11   )   &   -0.06   &   (   -0.01   ,   -0.11   )   \\  \hline
$B_u \to K^- K^0$       &   -0.01   &   (   0.94    ,   -0.94   )   &   0.00   &   (   0.79    ,   -0.79   )   \\  \hline
$B_d \to\pi^0 \pi^0$        &   0.60 &   (   0.77    ,   -0.85   )   &   0.53    &   (   0.67    ,   -0.17   )   \\  \hline
$B_d \to \bar K^0 K^0$      &   -0.01   &   (   0.94    ,   -0.94   )   &   0.00   &   (   0.79    ,   -0.79   )   \\  \hline
$B_d \to \pi^0 \bar K^0$    &   -0.12   &   (   0.15    ,   -0.18   )   &   -0.15   &   (   0.02    ,   -0.22   )   \\  \hline
$B_s \to K^+ \pi^-$         &   0.21    &   (   0.38    ,   0.03    )   &   0.15    &   (   0.27    ,   0.03    )   \\  \hline
$B_s \to K^0 \pi^0$         &   0.60 &   (   0.77    ,   -0.85   )   &   0.53    &   (   0.67    ,   -0.17   )   \\  \hline
$B_s \to K^+K^- $       &   -0.06   &   (   -0.01   ,   -0.11   )   &   -0.06   &   (   -0.01   ,   -0.11   )   \\  \hline
$B_s \to K^0 \bar K^0$      &   0.00   &   (   0.05    ,   -0.05   )   &   0.00   &   (   0.04    ,   -0.04   )   \\  \hline
\end{tabular}
\end{center}
\end{table}

\normalsize
\newpage
\section{Effects of annihilation contributions}

In the analyses of the previous sections we have
neglected annihilation contributions to
$B\to PP$ decays. In this section we study the effects
of the annihilation terms on $B\to PP$ decays.
The inclusion of annihilation contributions
introduce 6 more hadronic parameters. They are
\begin{eqnarray}
A^T_{\bar 3}e^{i\delta_{A^T_{\bar 3}}},
A^P_{\bar 3} e^{i\delta_{A^P_{\bar 3}}},
A^T_{\overline{15}} e^{i\delta_{A^T_{\overline{15}}}},
\end{eqnarray}
In total we would have 13 parameters. From Table \ref{br} we see that there are 15
experimental data points. In principle, the 13 hadronic parameters under consideration
can be determined. In Tables \ref{ann1}, \ref{ann3} and \ref{ann2} we show the results
on the hadronic parameters, $B\to PP$ branching ratios and CP asymmetries.

\normalsize
\begin{table}
\caption{The best fit values and their errors for the hadronic parameters with
annihilation terms and $\gamma = 59^\circ$.}\label{ann1}
\begin{center}
\begin{tabular}{|c|c|c|c|c|}
\hline
         &\multicolumn{2}{c}{SU(3) exact}\vline & \multicolumn{2}{c}{ SU(3) break}\vline  \\ \hline
            &   central value   &   error range &   central value   &   error range \\  \hline
$   C^P_{\bar 3}    $   &       0.138       &       0.004       &       0.113       &       0.003       \\  \hline
$   C^T_{\bar 3}    $   &       0.208       &       0.181       &       0.225       &       0.112       \\  \hline
$   C^T_6   $   &       0.043       &       0.206       &       0.095       &       0.163       \\  \hline
$   C^T_{\overline{15}} $   &       0.141       &       0.014       &       0.143       &       0.014       \\  \hline
$   \delta_{  \bar 3}   $   &   $   31.65   ^0$ &   $   57.7    ^0$ &   $   26.42  ^0$ &   $   35.94  ^0$ \\  \hline
$   \delta_ 6           $   &   $   97.74   ^0$ &   $   147.97  ^0$ &   $   81.90  ^0$ &   $   52.38   ^0$ \\  \hline
$   \delta_{\overline{15}}  $   &   $   8.54    ^0$ &   $   21.67   ^0$ &   $   4.58   ^0$ &   $   16.75  ^0$ \\  \hline  \hline
$   A^P_{\bar 3}    $   &       0.025       &       0.042       &       0.017       &       0.022       \\  \hline
$   A^T_{\bar 3}    $   &       0.061       &       0.143       &       0.039       &       0.082       \\  \hline
$   A^T_{\overline {15}}    $   &       0.036       &       0.075       &       0.023       &       0.043       \\  \hline
$   \delta_{A^P_{\bar{3}}}  $   &   $   -13.78  ^0$ &   $   89.52   ^0$ &   $   -42.27  ^0$ &   $   98.46   ^0$ \\  \hline
$   \delta_{A^T_{\bar{3}}}  $   &   $   73.46   ^0$ &   $   107.18  ^0$ &   $   54.88   ^0$ &   $   113.5   ^0$ \\  \hline
$   \delta_{A^T_{\overline{15}}}    $   &   $   -131.12  ^0$ &   $   180.51  ^0$ &   $   -175.94 ^0$ &   $   197.56  ^0$ \\  \hline
\end{tabular}
\end{center}
\end{table}

\footnotesize
\begin{table}
\caption{The prediction of the branching ratios with annihilation
terms and $\gamma$=$59^{\circ}$}\label{ann3}
\begin{center}
\begin{tabular}{|c|c|c|c|c|}
Branching ratio &   \multicolumn{2}{c}{ SU(3) Exact}\vline                          &   \multicolumn{2}{c}{ SU(3) break}\vline                          \\  \hline
    &   central value   &   Error(  Max ,   Min )   &   central value   &   Error(  Max ,   Min )   \\  \hline
$B_u \to \pi^- \pi^0$       &   5.2 &   (   6.2 ,   4.1 )   &   5.3 &   (   6.4 ,   4.3 )   \\  \hline
$B_u \to K^- K^0$       &   0.7 &   (   1.0   ,   0.6 )   &   1.0   &   (   1.3 ,   0.9 )   \\  \hline
$B_d \to \pi^+ \pi^-$       &   5.2 &   (   5.8 ,   4.5 )   &   5.1 &   (   5.7 ,   4.5 )   \\  \hline
$B_d \to\pi^0 \pi^0$        &   1.4 &   (   2.2 ,   0.7 )   &   1.2 &   (   2.0   ,   0.6 )   \\  \hline
$B_d \to K^- K^+$       &   0.1 &   (   0.4 ,   0.0   )   &   0.1 &   (   0.4 ,   0.0   )   \\  \hline
$B_d \to \bar K^0 K^0$      &   1.9 &   (   4.4 ,   0.4 )   &   2.2 &   (   4.5 ,   0.8 )   \\  \hline
$B_u \to \pi^- \bar K^0$    &   18.9    &   (   20.2    ,   17.5    )   &   18.8    &   (   20.0  ,   17.6    )   \\  \hline
$B_u \to \pi^0 K^-$         &   10.3    &   (   11.1    ,   9.6 )   &   10.6    &   (   11.2    ,   10.0  )   \\  \hline
$B_d \to \pi^+ K^-$         &   18.6    &   (   19.7    ,   17.6    )   &   18.6    &   (   19.5    ,   17.6    )   \\  \hline
$B_d \to \pi^0 \bar K^0$    &   8.6 &   (   9.6 ,   8.1 )   &   8.4 &   (   8.9 ,   7.9 )   \\  \hline  \hline
$B_s \to K^+ \pi^-$         &   4.1 &   (   6.7 ,   2.6 )   &   6.4 &   (   8.8 ,   4.7 )   \\  \hline
$B_s \to K^0 \pi^0$         &   1.1 &   (   2.2 ,   0.4 )   &   1.4 &   (   2.6 ,   0.6 )   \\  \hline
$B_s \to \pi^+\pi^- $       &   2.3 &   (   9.5 ,   0.0   )   &   1.1 &   (   4.3 ,   0.0   )   \\  \hline
$B_s \to \pi^0 \pi^0$       &   1.1 &   (   4.7 ,   0.0   )   &   0.5 &   (   2.2 ,   0.0   )   \\  \hline
$B_s \to K^+K^- $       &   31.8    &   (   51.9    ,   7.1 )   &   41.3    &   (   66.6    ,   14.8    )   \\  \hline
$B_s \to K^0 \bar K^0$      &   30.9    &   (   50.7    ,   6.5 )   &   39.6    &   (   65.0  ,   13.5    )   \\  \hline
\end{tabular}
\end{center}
\end{table}

\scriptsize
\begin{table}
\caption{The prediction of the CP asymmetry with
annihilation terms and $\gamma$=$59^{\circ}$}\label{ann2}
\begin{center}
\begin{tabular}{|c|c|c|c|c|}
\hline
Asymmetry   &   \multicolumn{2}{c}{ SU(3) Exact}\vline                          &   \multicolumn{2}{c}{ SU(3) break}\vline                          \\  \hline
    &   Central value   &       Error (Max,Min)             &   Central value   &       Error (Max,Min)             \\  \hline
$B_u \to \pi^- \pi^0$       &   0.00   &   (   0.00   ,   0.00   )   &   0.00   &   (   0.00   ,   0.00   )   \\  \hline
$B_d \to \pi^+ \pi^-$       &   0.41    &   (   0.61    ,   0.20 )   &   0.37    &   (   0.52    ,   0.18    )   \\  \hline
$B_u \to \pi^- \bar K^0$    &   0.01    &   (   0.05    ,   -0.04   )   &   0.00   &   (   0.04    ,   -0.03   )   \\  \hline
$B_u \to \pi^0 K^-$         &   -0.03   &   (   0.05    ,   -0.14   )   &   -0.02   &   (   0.06    ,   -0.12   )   \\  \hline
$B_d \to \pi^+ K^-$         &   -0.06   &   (   -0.01   ,   -0.11   )   &   -0.07   &   (   -0.02   ,   -0.12   )   \\  \hline  \hline
$B_u \to K^- K^0$       &   -0.24   &   (   0.82    ,   -0.96   )   &   -0.08   &   (   0.73    ,   -0.81   )   \\  \hline
$B_d \to\pi^0 \pi^0$        &   0.19    &   (   0.72    ,   -0.99   )   &   0.19    &   (   0.62    ,   -0.86   )   \\  \hline
$B_d \to K^- K^+$       &   0.83    &   (   1.00   ,   -1.00  )   &   0.77    &   (   1.00   ,   -1.00  )   \\  \hline
$B_d \to \bar K^0 K^0$      &   0.78    &   (   1.00   ,   -1.00  )   &   0.40 &   (   1.00   ,   -1.00  )   \\  \hline
$B_d \to \pi^0 \bar K^0$    &   -0.02   &   (   0.13    ,   -0.12   )   &   -0.04   &   (   0.10 ,   -0.13   )   \\  \hline
$B_s \to K^+ \pi^-$         &   0.26    &   (   0.54    ,   0.03    )   &   0.20 &   (   0.36    ,   0.06    )   \\  \hline
$B_s \to K^0 \pi^0$         &   0.12    &   (   0.68    ,   -0.86   )   &   0.21    &   (   0.62    ,   -0.64   )   \\  \hline
$B_s \to \pi^+\pi^- $       &   -0.04   &   (   1.00   ,   -1.00  )   &   -0.04   &   (   1.00   ,   -1.00  )   \\  \hline
$B_s \to \pi^0 \pi^0$       &   -0.04   &   (   1.00   ,   -1.00  )   &   -0.04   &   (   1.00   ,   -1.00  )   \\  \hline
$B_s \to K^+K^- $           &   -0.06   &   (   -0.02   ,   -0.24   )   &   -0.10    &   (   -0.04   ,   -0.23   )   \\  \hline
$B_s \to K^0 \bar K^0$      &   -0.05   &   (   0.18    ,   -0.22   )   &   -0.02   &   (   0.11    ,   -0.14   )   \\  \hline
\end{tabular}
\end{center}
\end{table}

\normalsize

From Table \ref{ann1} we see that the size of
the best fit annihilation parameters $A_i$ are small
compared with the  non-annihilation terms $C_{\bar 3,\overline{15}}$.
This confirms the conjecture that annihilation
contributions are small.
The allowed ranges are, however, large and therefore can not
rule out the possibility of having significant annihilation contributions.
We have to wait improved experiments to obtain more precise information.
We note that $A_i$ actually have similar size as $C^T_6$.

The branching ratios for  $B_d \to K^- K^+$, $B_s \to \pi^+\pi^-$ and $B_s \to
\pi^0\pi^0$ which only receive contribution from annihilation are not vanishing any more.
The branching ratios are expected to be small. From Table \ref{ann3},
we indeed find that these branching ratios are among the small ones.

It is interesting to note that although the
annihilation amplitudes are small, in certain decay modes, such as
$B_s \to K^+K^-$ and $B_s \to K^0 \bar
K^0$, the effects on the branching ratios can be significant.
This is because that although  $A^P_{\bar 3}$ is small
compared with $C_{\bar 3,\overline{15}}$, and is comparable with $C^T_{6}$, but
enhanced by a KM factor $|V_{tb}V_{ts}^*/V_{ub}V_{us}^*|$.
These modes provide good places to
study the annihilation contributions. It can be seen that
SU(3) breaking effects are also large in these decays. From
Table \ref{ann2}, we also see that CP violation can be affected significantly.
CP asymmetries in $B_d \to K^0\bar K^0 $ can be more than 50\% with a not
too small branching ratio.

\section{Discussions and Conclusions}

We have studied branching ratios and CP violating rate asymmetries
in $B\to PP$ decays in the Standard Model using SU(3) flavor
symmetry. In the SM when annihilation
contributions are neglected only seven hadronic
parameters are needed to describe $B\to PP$ decays,
six more hadronic parameters are needed to include
the annihilation contribution.
We have shown that present experimental data on these
decays can be used to systematically determine hadronic
parameters, in particular the CP conserving FSI phases.

Although great efforts have been made to understand the
dynamics of low energy strong  interactions to calculate
theoretically the decay amplitudes and the CP conserving FSI
phases for $B\to PP$ decays, such as factorization approximation
with improvement from QCD corrections\cite{BBNS}. It is still far away from
being able to predict with high confidence level the amplitudes.
Still factorization calculations may provide some ideal
about the order of magnitude.
We have numerically studied the predictions of
factorization approximation for the size of the SU(3) invariant amplitudes.
We found that the size of the hadronic amplitudes
obtained in this paper are in the same order of magnitudes
as those from factorization calculations\cite{6b},
but the FSI phases, which can not be reliably
calculated in factorization approximation, can be
very different and large. We also found that the
annihilation contributions are generally small, but can have
significant effects on some decays,
such as $B_s \to K^+ K^-, K^0 \bar K^0$.

We attempted to study SU(3) breaking effects
in $B\to PP$ decays by assuming a simple pattern for
the breaking effects. We found that although the general
features are not changed very much, in certain decays the
effects can be large, such as the branching ratios
for $B_s \to K^+ K^-, K^0 \bar K^0$. Therefore these modes can
be good modes to study SU(3) breaking effects.

We predicted branching ratios for several $B_s \to PP$ decays.
These decay branching ratios can be measured at future hadron colliders.
The SM and SU(3) flavor symmetry can be tested.

At present CP violating rate asymmetries in $B\to PP$ have not been
measured. The use of SU(3) flavor symmetry can also provide important
information on CP violation in the Standard Model. Using the best
fit values for the hadronic parameters, we also obtained CP violating rate
asymmetries for $B\to PP$ decays.
We found that some of the asymmetries can be large and within the reach of $B$
factories. CP asymmetry in $B_d \to \pi^+\pi^-$ can be as large as
30\% and even larger ones for $B_d \to K K$.
It can be expected
that with more accurate experimental measurements, the study of
CP violating rate asymmetries can provide crucial information
about dynamics for $B$ decays in the Standard Model.


\begin{references}
\bibitem{1}
M. Savage and M. Wise, Phys. Rev.
{\bf D39}, 3346(1989); $ibid$ {\bf D40}, Erratum, 3127(1989);
X.-G. He, Eur. Phys. J. {\bf C9}, 443(1999);
N. G. Deshpande, X.-G. He, and J.-Q. Shi, Phys. Rev. {\bf D62}, 034018(2000).


\bibitem{2} M. Gronau et al., Phys. Rev. {\bf D50}, 4529 (1994); {\bf D52},
6356 (1995); $ibid$, 6374 (1995);
A.S. Dighe, M. Gronau and J. Rosner, Phys. Rev. Lett. {\bf 79}, 4333 (1997); L.L. Chau et al.,
Phys. Rev. {\bf D43}, 2176 (1991); D. Zeppendfeld, Z. Phys. {\bf C8}, 77(1981).

\bibitem{3} M. Gronau and D. London, Phys. Rev. Lett. {\bf 65},
3381(1990);
M. Gronau and J. Rosner,
Phys. Rev. {\bf D57}, 6843(1998); $ibid$, {\bf D61}, 073008(2000);
Phys. Lett. {\bf B482}, 71(2000), M. Gronau, D. Pirjol and T.-M. Yan,
Phys. Rev. {\bf D60}, 034021(1999).

\bibitem{4} X.-G. He, J.-Y. Leou and C.-Y. Wu, Phys. Rev. {\bf D62},
114015(2000); N. G. Deshpande and X.-G. He,
Phys. Rev. Lett. {\bf 75}, 1703(1995).

\bibitem{5} Y.-F. Zhou et al., Phys. Rev. {\bf D63}, 054011(2001);
M. Bargiotti et al., e-print hep-ph/0204029.


\bibitem{6a} R. Fleischer and T. Mannel, Phys. Rev. {\bf D57}, 2752(1998);
M. Neubert and J. Rosner, Phys. Lett. {\bf B441}, 403(1998);
M. Neubert and J. Rosner, Phys. Rev. Lett. {\bf 81}, 5076(1998);
X.-G. He, C.-L. Hsueh and J.-Q. Shi, Phys. Rev. Lett.
{\bf 84} 18(2000); M. Gronau and J. Rosner, Phys. Rev. {\bf D57}, 6843(1998);
N.G. Deshpande and X.-G. He, Phys. Rev. Lett. {\bf 75}, 3064(1995).

\bibitem{6b}  X.-G. He et al. ,Phys.Rev. {\bf D64} 034002(2001).



\bibitem{8} Particle Data Group, Eur. Phys. J. {\bf C 15} 1(2000).


\bibitem{10}  S. Mele, Phys.Rev. {\bf D59} 113011(1999);
A. Ali, D. London, Eur.Phys.J. {\bf C9} 687(1999); F. Parodi, P. Roudeau and A. Stocchi, Nuovo. Cim. {\bf A112}, 833(1999);
F. Caravaglios et al., e-print hep-ph/0002171.


\bibitem{ampmd} H. G. Moser and A. Roussani, Nucl. Inst. Meth. {\bf A384},491(1997).


\bibitem{12} T. Affolder et al. (CDF Collaboration), Phys. Rev. {\bf D61}, 072205(2000).
\bibitem{13} The ALEPH Collaboration, Phys. Lett. {\bf 492}, 259(2000).

\bibitem{14} The Babar Collaboration, e-print hep-ex/0203007.

\bibitem{15} The Belle Collaboration, e-print hep-ex/0205020.

\bibitem{19} http://lepbosc.web.cern.ch/LEPBOSC, CKM workshop February 2002, CERN, Geneva.

\bibitem{20} G.Boix, e-print hep-ex/0104048.

\bibitem{dh} N. Deshpande and X.-G. He, Phys. Lett. {\bf B336}, 471(1994).

\bibitem{21}
D. Cronin-Hennessy, et al. (CLEO Collaboration), Phys. Rev. Lett. {\bf 85}, 515(2000);
D.M. Asner, et al. (CLEO Collaboration), Phys. Rev. {\bf D65}, 031103(2002).
\bibitem{22}
M.Z.Wang (Belle Collaboration),  talk presented at
Lake Louise Winter Institute 2002 on Fundamental Interactions,
Lake Louise, Alberta, Canada, February (2002).
\bibitem{23}
B. Aubert, et al. (Babar Collaboration), Phys. Rev. Lett. {\bf 87}, 151802 (2001).
J. Olsen and M. Bona (Babar Collaboration), talks presented at
American Physical Society's 2002 Meeting of The Division of Particles and Field,
Williamsburg, Virginia, USA, May 24 - 28 (2002).

\bibitem{js} C. Jarlskog, Phys. Rev. Lett. 55, 1093 (1985); Z.
Phys. C29, 491 (1985); O.W. Greenberh, Phys, Rev. D 32, 1841
(1985); D.-D. Wu, Phys, Rev. D 33,860 (1986).

\bibitem{BBNS} M.~Beneke et al.,
Phys. Rev. Lett. {\bf 83}, 1914 {1999}; Nucl. Phys. {\bf B591}, 313 (2000);
Y.-Y. Keum, H.-n. Li and I. Sanda, Phys. Lett. {\bf B504},
6(2001); Phys. Rev. {\bf D63}, 054008(2001).

\end{references}
\end{document}